\documentclass[acmsmall]{acmart}

\AtBeginDocument{%
  }

\setcopyright{acmlicensed}
\copyrightyear{2018}
\acmYear{2018}
\acmDOI{XXXXXXX.XXXXXXX}

\acmConference[Conference acronym 'XX]{Make sure to enter the correct
  conference title from your rights confirmation emai}{June 03--05,
  2018}{Woodstock, NY}
\acmISBN{978-1-4503-XXXX-X/18/06}

\usepackage[algoruled,ruled,vlined,lined,commentsnumbered]{algorithm2e}

\usepackage{colortbl}
\usepackage{tabularx}
\usepackage{multirow}
\usepackage{multicol}
\usepackage{rotating}
\usepackage{wrapfig}
\usepackage{xspace}
\usepackage{tcolorbox}
\usepackage{subcaption}

\newcounter{myboxcounter}
\definecolor{grey}{RGB}{224,224,224}
\definecolor{tableblue}{RGB}{220,234,247}
\definecolor{myblue}{RGB}{166,202,236}
\definecolor{myorange}{RGB}{255,225,185}
\definecolor{myred}{RGB}{246,198,173}
\definecolor{tbred}{RGB}{255,205,205}
\definecolor{tbgreen}{RGB}{217,242,208}
\definecolor{rqred}{RGB}{249,185,188}
\definecolor{rqyellow}{RGB}{255,230,153}

\newcommand{\autofl}{AutoFL\xspace}
\newcommand{\agentfl}{AgentFL\xspace}

\newcommand{\toolname}{$\mathtt{CosFL}$\xspace}
\newcommand{\bftoolname}{$\mathbf{\mathtt{CosFL}}$\xspace}
\newcommand*{\eg}{e.g., }
\newcommand*{\ie}{i.e., }
\newcommand*{\etc}{etc. }
\newcommand{\etal}{\emph{et~al.}\xspace}

\newcommand{\prompt}[3]{
\refstepcounter{myboxcounter}
\begin{tcolorbox}[colback=gray!10, colframe=black!80,
width=\linewidth, arc=2mm, auto outer arc, title={{\small #1}}, label={prompt:#2}, center]
{
\begingroup
\small
\linespread{0.8}\selectfont
#3
\endgroup
}
\end{tcolorbox}
}

\newcommand{\notez}[1]{
\begin{tcolorbox}[size=fbox,boxrule=0.5pt,top=0.5pt,bottom=0.5pt,
colframe=blue!5!black,colback=black!5!white]
\em #1
\end{tcolorbox}
}

\begin{document}

\title{Fault Localization from the Semantic Code Search Perspective}

\author{Yihao Qin}
\email{yihaoqin@nudt.edu.cn}
\affiliation{%
  \institution{National University of Defense Technology}
  \city{Changsha}
  \state{Hunan}
  \country{China}
}

\author{Shangwen Wang}
\authornote{Corresponding author.}
\affiliation{%
  \institution{National University of Defense Technology}
  \city{Changsha}
  \state{Hunan}
  \country{China}
}
\email{wangshangwen13@nudt.edu.cn}

\author{Yan Lei}
\affiliation{%
  \institution{Chongqing University}
  \city{Chongqing}
  \country{China}
}
\email{yanlei@cqu.edu.cn}

\author{Zhuo Zhang}
\affiliation{%
 \institution{Tianjin University}
 \city{Tianjin}
 \country{China}
}
\email{zz8477@126.com}

\author{Bo Lin}
\affiliation{%
  \institution{National University of Defense Technology}
  \city{Changsha}
  \state{Hunan}
  \country{China}
}
\email{linbo19@nudt.edu.cn}

\author{Xin Peng}
\affiliation{%
  \institution{National University of Defense Technology}
  \city{Changsha}
  \state{Hunan}
  \country{China}
}
\email{xinpeng@nudt.edu.cn}

\author{Liqian Chen}
\affiliation{%
  \institution{National University of Defense Technology}
  \city{Changsha}
  \state{Hunan}
  \country{China}
}
\email{lqchen@nudt.edu.cn}

\author{Xiaoguang Mao}
\affiliation{%
  \institution{National University of Defense Technology}
  \city{Changsha}
  \state{Hunan}
  \country{China}
}
\email{xgmao@nudt.edu.cn}

\renewcommand{\shortauthors}{Trovato et al.}

\begin{abstract}
The software development process is characterized by an iterative cycle of continuous functionality implementation and debugging, essential for the enhancement of software quality and adaptability to changing requirements.
This process incorporates two isolatedly studied tasks: Code Search (CS), which retrieves reference code from a code corpus to aid in code implementation, and Fault Localization (FL), which identifies code entities responsible for bugs within the software project to boost software debugging.
The basic observation of this study is that these two tasks exhibit similarities since they both address search problems. Notably, CS techniques have demonstrated greater effectiveness than FL ones, possibly because of the precise semantic details of the required code offered by natural language queries, which are not readily accessible to FL methods. 
Drawing inspiration from this, we hypothesize that a fault localizer could achieve greater proficiency if semantic information about the buggy methods were made available.
Based on this idea, we propose \toolname, an FL approach that decomposes the FL task into two steps: \emph{query generation}, which describes the functionality of the problematic code in natural language, and \emph{fault retrieval}, which uses CS to find program elements semantically related to the query, allowing for finishing the FL task from a CS perspective.
Specifically, to depict the buggy functionalities and generate high-quality queries, \toolname extensively harnesses the code analysis, semantic comprehension, text generation, and decision-making capabilities of LLMs. 
Moreover, to enhance the accuracy of CS, \toolname captures varying levels of context information and employs a multi-granularity code search strategy, which facilitates a more precise identification of buggy methods from a holistic view.
The evaluation on 835 real bugs from 23 Java projects shows that \toolname successfully localizes 324 bugs within Top-1, which significantly outperforms the state-of-the-art approaches by 26.6\%-57.3\%. The ablation study and sensitivity analysis further validate the importance of different components and the robustness of \toolname across different backend models.
\end{abstract}


\received{20 February 2007}
\received[revised]{12 March 2009}
\received[accepted]{5 June 2009}

\maketitle
\section{Introduction}

The software development process is an iterative cycle of continuous functionality implementation and debugging~\cite{katz1987debugging}.
During the implementation phase, programmers write code that meets specific requirements, but may simultaneously introduce bugs into the software system.
In the debugging phase, developers search for the root causes through error signals and fix those program behaviors that do not meet expectations.
This iterative process is crucial for continuously improving software quality and flexibly adapting to changing requirements.
In the literature, two types of techniques are widely studied to facilitate development activities in this cycle, one for each phase.
The first is {\bf Code Search} (CS)~\cite{sun2024survey} during the implementation phase, which aims to assist developers in reusing specific code snippets from open source repositories, rather than ``reinventing the wheel''.
The second is {\bf Fault Localization} (FL)~\cite{wong2023survey} during the debugging phase, which aims to identify the buggy program elements within the entire software system, and thus speed up the process of combating the bugs.

During the years, these two types of techniques have generally been studied in isolation, with few existing studies attempting to associate them. 
The only obvious commonality we can identify between these two types of techniques is the trend for both to leverage the strengths of advanced deep learning techniques in their technical development.
Specifically, owing to the advancements in representation learning~\cite{wan2024survey}, CS techniques have evolved from traditional keyword matching to learning the semantic correlation between the requirements and code snippets.
Similarly, FL techniques have also evolved from coverage-based program spectrum~\cite{abreu2007jaccard} to representation learning, or even empowered by latest Large Language Models (LLMs)~\cite{kang2024autofl}.
Despite their shared trend, there is a significant discrepancy in the effectiveness of these two types of techniques.
Experiments conducted on domain-specific benchmarks have shown that state-of-the-art CS techniques~\cite{wang2023codesearch} can accurately identify the required code snippets for approximately 80\% of the total requirements. In contrast, the latest FL technique~\cite{kang2024autofl} can only locate the buggy program entities at the top position for around 30\% of the cases.

\begin{figure}[t]
  \centering
  \includegraphics[width=0.7\linewidth]{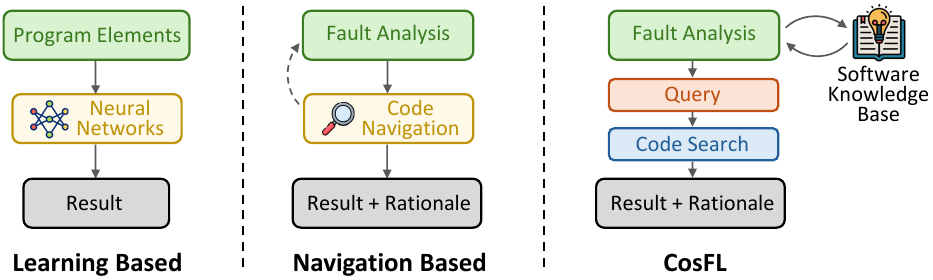}
  \caption{Illustration of \toolname and other FL pipelines. The rationale is generated by LLMs.}
  \label{fig:intro}
  \Description{}
\end{figure}

To boost the effectiveness of FL, the basic observation of this study is that the tasks of code search and fault localization inherently share certain similarities, as they both solve a search problem.
In practical terms, CS takes requirements described in natural language queries as input and retrieves code from a large repository that aligns with these intentions. In this context, the queries contain rich semantic information about the required functionalities, allowing for the accurate identification of the necessary code~\cite{shi2023cocosoda}. 
On the other hand, FL takes a buggy project as input and identifies the program entities responsible for the bug. While FL has access to readily available information such as the source code, test suite, and failure information, it lacks explicit semantic information about the functionality of the buggy code.
To address this shortfall, researchers have delved into leveraging advanced deep learning techniques, including coverage-based graph representation learning~\cite{lou2021grace} and function call-based code navigation~\cite{kang2024autofl}. Despite these efforts, there remains ample room for enhancements in the efficacy of current fault localization techniques as aforementioned.
Considering the differences in performance and inputs between these two types of techniques, we postulate that a fault localizer may perform better if it is provided with semantic information about the functionality of the buggy code.
Consequently, the key idea of this study is to {\bf equip the fault localizer with a natural language query that elucidates the functionality of the erroneous code, enabling fault localization to be performed in a code search fashion}.

Based on this idea, we propose \toolname, a \textbf{CO}ode \textbf{S}earch inspired \textbf{F}ault \textbf{L}ocalization approach that localizes bugs at the method level.
As illustrated in Figure \ref{fig:intro}, \toolname decomposes the fault localization task into two steps, namely, {\em query generation} and {\em fault retrieval}.
In the first step, a natural language query is generated to describe the functionality of the possible problematic code, while in the second step, CS is employed to match suspicious program elements semantically related to the query from the entire project.
As shown, this pipeline differs significantly from the workflows of existing representation learning-based FL~\cite{lou2021grace} and code navigation-based FL~\cite{kang2024autofl}, thus exploring the feasibility of a new fashion for FL.
Intuitively, our pipeline faces two primary challenges: first, generating high-quality queries, as this information is not readily available and requires deduction through intricate code logic; 
and second, precisely pinpointing the buggy code, as the accuracy of code search still encounters challenges, particularly when different code segments within the same project exhibit semantic overlaps.
To address the first challenge, we leverage the capabilities of LLMs for their promising abilities towards code analysis~\cite{wu2023llmfl}, semantic understanding~\cite{nam2024understanding}, text generation~\cite{ahmed2023summarization}, as well as autonomous decision-making~\cite{jin2024llms}. 
In particular, we first construct a software knowledge base for the target project by amalgamating static analysis details and dynamic method call graphs. The LLM autonomously determines whether the contextual information suffices to describe the buggy functionality; if not, we supplement it with essential project-specific knowledge from the base. This iterative process unfolds through multi-turn interactions and terminates when the LLM attains adequate confidence to produce a high-quality query.
To tackle the second challenge, we implement a multi-granularity code search strategy that conducts retrieval at various program entity levels (\ie chunk, method, and module), allowing for a more comprehensive depiction of the semantics across different code snippets.
Finally, a voting mechanism from a holistic perspective is employed to pinpoint the fault localization results.

We comprehensively evaluated the effectiveness of \toolname on Defects4J~\cite{just2014defects4j} and GrowingBugs~\cite{jiang2022growingbugs} benchmarks, which consist of 835 real bugs from 23 Java projects.
The results indicate that \toolname performs well by successfully pinpointing the fault location within the Top-1 position for 38.8\% (324 out of 835) of the bugs.
Compared to other code navigation-based approaches, \toolname achieves a significantly higher localization accuracy, surpassing \agentfl~\cite{qin2024agentfl} and \autofl~\cite{kang2024autofl} on the Top-1 metric by 26.6\% and 57.3\%, respectively.
We further validated the importance of different components in \toolname with an ablation study.
Additionally, we conducted a sensitivity analysis to illustrate the impact of different parameter settings on \toolname.
The results demonstrate that \toolname maintains robustness across various LLMs and embedding models. 
Finally, we presented specific case studies to demonstrate how \toolname performs better than other approaches in FL tasks.

Overall, the main contributions of this paper include:
\begin{itemize}
    \item \textbf{Perspective.} To boost the performance of FL, we propose to conduct this task from a new perspective, \ie semantic code search.
    \item \textbf{Methodology.} Based on our idea, we propose \toolname, a semantic code search-based fault localizer supported by LLMs' strong code understanding and reasoning capabilities.
    \item \textbf{Experiment.} We comprehensively evaluate \toolname on 835 bugs from 23 projects, the results demonstrate that \toolname significantly outperforms existing LLM-empowered FL techniques.
\end{itemize}
\section{Motivation}
\label{sec:motiv}

\begin{figure}[t]
  \centering
  \includegraphics[width=0.9\linewidth]{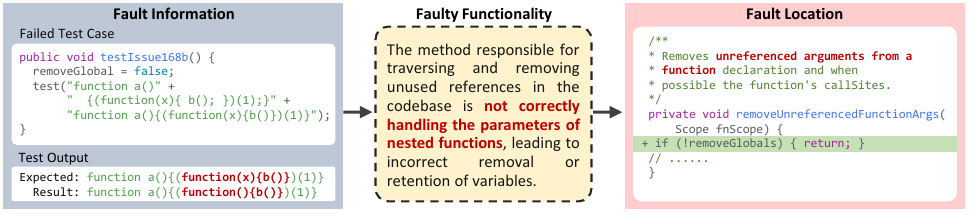}
  \caption{Concept diagram of localizing Closure-1 bug from the perspective of semantic code search.}
  \label{fig:motiv}
  \Description{}
\end{figure}

Figure \ref{fig:motiv} illustrates a bug from the closure-compiler~\footnote{\url{https://storage.googleapis.com/google-code-archive/v2/code.google.com/closure-compiler/issues/issue-253.json}} project.
The root cause of this bug is that the function arguments were optimized away under unexpected situations.
In the \emph{Failed Test Case} \texttt{testIssue168b}, the expected program behavior is to parse the nested function \texttt{(function(x){b();})} as \texttt{(function(x){b()})}. However, in the \emph{Test Output}, the argument \texttt{x} has been accidentally removed. To fix this bug, the developers added control logic to the \texttt{removeUnreferencedFunctionArgs} method, preventing the program from deleting arguments when the variable \texttt{removeGlobal} is set to false.
As stated in the code comment, the changed method is responsible for \emph{``Removing unreferenced arguments from a function declaration and when possible the function's call sites''}.

From the perspective of semantic code search, the new FL paradigm envisioned in this work can be broadly divided into two steps:
First, given the \emph{Fault Information} in Figure \ref{fig:motiv}, we comprehensively analyze the root cause of the fault to generate the \emph{Faulty Functionality}, which describes the possible root cause in natural language as \emph{``not correctly handling the parameters of nested function''}.
Such a description can indicate the basic semantic information of the buggy function (\eg we can infer that the buggy function handles the parameters of nested function in this case).
Subsequently, we transform the process of searching for suspicious program elements in FL into a CS task, which takes the \emph{Faulty Functionality} as a query to match program elements that are semantically similar to it within the entire codebase.
In this case, the approach we proposed successfully identifies the buggy method as the most suspicious location.
Intuitively, enhancing FL with CS offers numerous advantages over traditional code navigation-based FL.
Code search reviews the entire codebase directly, rather than progressively narrowing the search scope from coarse to fine-grained, which may miss the fault location due to inaccurate decisions.
To help further understand the superiority of the new pipeline, we have demonstrated more case studies in Sec \ref{subsec:case}.


However, despite the extensive research on CS and FL over the past decades, there has been a lack of studies applying the spirit of CS to FL. We attribute this to the absence of prerequisite technologies.
Specifically, for semantic-based FL, a significant gap exists between fault-inducing phenomena (such as program output, error stack trace, and failing test case) and fault functionality descriptions that can be used for semantic code search.
Traditional statistical learning methods and representation learning approaches struggle to bridge this gap.
Fortunately, with the continuous advancement of artificial intelligence technologies, the emergence of large language models has made it possible to automatically diagnose problems in programs~\cite{wu2023llmfl,openai2024gpt4}, thus filling the gap in FL from the CS perspective.
In the next section, we will explain our proposed solution in detail.

\section{Approach}
\label{sec:approach}

In this work, we propose \toolname, a new paradigm for \textbf{F}ault \textbf{L}ocalization from the perspective of semantic \textbf{CO}de \textbf{S}earch.
\toolname primarily focuses on method-level FL, since this granularity level is most preferred by developers~\cite{kochhar2016practitioner} and also serves as the foundation for many automated software development techniques~\cite{yang2024swe}.
To achieve better compatibility between CS and FL, \toolname provides two solutions:
1) To generate high-quality queries, \toolname constructs a software knowledge base, providing useful context for LLM to understand the root causes comprehensively.
2) To improve the search precision, we implement multi-granularity semantic CS to more accurately localize buggy methods, and further optimize the localization results through a voting mechanism.

\begin{figure}[t]
  \centering
  \includegraphics[width=1.0\linewidth]{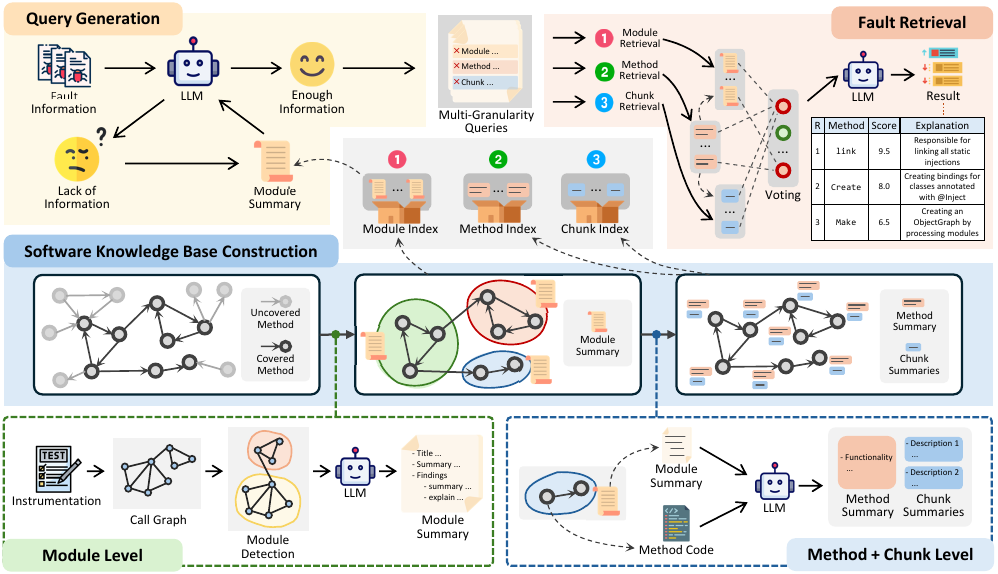}
  \caption{Overview of \toolname.}
  \label{fig:approach}
  \Description{The Overview of \toolname.}
\end{figure}

An overview of \toolname is shown in Figure \ref{fig:approach}.
In this work, we define a \emph{module}~\cite{koru2009module} as a group of closely functionally related methods and a \emph{chunk}~\cite{sobreira2018dissection} as a sequence of continuous statements in a method.
These two levels of granularity, describing program semantics from both a holistic and a detailed perspective, respectively, contribute to comprehensively depicting the context of methods and precisely pinpointing the fault locations.
In the \colorbox{myblue}{Software Knowledge Base Construction} phase, we utilize program instrumentation to collect the runtime behavior of the software, and then extract software knowledge at different granularities (i.e., module, method, and chunk) through LLM.
In the \colorbox{myorange}{Query Generation} phase, LLM starts with the fault information to generate queries that describe the suspicious software functionalities in multiple granularities.
When LLM is confused, we allow it to actively request module-level knowledge to more comprehensively understand the software's internal behavior.
In the \colorbox{myred}{Fault Retrieval} phase, \toolname employs CS to search for program elements at different granularities, and finally output a list of suspicious methods through a voting mechanism.

\subsection{Software Knowledge Base Construction}
\label{subsec:knowledge}

Given software with bugs, \toolname first collects the runtime behavior of the software and uses LLM to extract project-specific knowledge at different granularities. The acquired multi-granularity knowledge 
will provide informational support for the subsequent FL process. 
Specifically, we construct the software knowledge base for addressing two limitations:
1) From the fault analysis perspective, existing LLM-based FL methods~\cite{kang2024autofl} only provide basic fault information as input, such as test case and test output, while lacking a deep understanding of the complete software architecture and runtime behavior.
This may threaten LLM's judgment and consequently affect the FL effectiveness;
2) From the code retrieval perspective, semantically similar code snippets may exist at different locations in one project, necessitating fine-grained context information (\eg chunk-level semantics) for effective differentiation.
As shown in the lower part of Figure \ref{fig:approach}, the knowledge base construction is divided into two stages.
At a coarse granularity, \toolname first collects dynamic runtime information of the software and extracts module-level knowledge.
Based on the module-level knowledge, \toolname further extracts method- and chunk-level knowledge for each method at a fine granularity.
In the following content, we will introduce these two stages in detail.

\textbf{Module-Level Knowledge Extraction.}
At this stage, our objective is to acquire module-level knowledge within the software.
According to the consensus in software architecture~\cite{koru2009module}, we naturally consider methods as minimal functional units and define a module as a group of closely functionally related methods.
Specifically, to more accurately represent the software behavior that triggers the bug, the functional modules mentioned in this paper are dynamically constructed, that is, all methods comprising a module must have been executed at runtime.

\emph{Call Graph Construction.}
To start with, we employ program instrumentation techniques to collect method invocation logs during software execution and generate a dynamic method call graph on the fly.
In this graph, we record the method ID, invocation edges, and the frequency of each invocation.
During this process, we exclude methods that are not covered during runtime, thereby avoiding the extraction of knowledge contaminated by untriggered program functionalities.

\emph{Module Detection.}
However, in large-scale software systems, dynamic method call graphs can be extremely extensive.
For instance, the call graph of the example mentioned in Figure \ref{fig:motiv} contains 1,082 nodes and 2,064 edges.
Given that LLMs struggle to comprehend inputs of such magnitude~\cite{liu2024lost}, it is necessary to appropriately decompose the entire call graph (corresponding to the overall software behavior) into smaller subgraphs (corresponding to the functional modules).
To this end, we noticed that the principle of defining functional modules in software engineering is high cohesion and low coupling~\cite{hitz1995measuring}, a characteristic highly similar to communities in networks.
In graph theory, a community can be defined as a subset of nodes that are densely connected to each other but sparsely connected to nodes in other communities within the same graph~\cite{bedi2016community}.
Inspired by this concept, we identify functional modules within the call graph through the Leiden~\cite{traag2019leiden} community detection algorithm.
The Leiden algorithm optimizes modularity and identifies tightly connected groups within the network through an iterative process of moving nodes to neighboring communities, refining the community structure, and then aggregating the communities into new nodes.
In the algorithm, the quality function serves as the optimization objective, guiding node movement and community partitioning, evaluating the quality of divisions, and affecting the convergence and outcomes of the algorithm.
In this work, we measure the quality of a detected module by the density and frequency of method invocations within it.
Our intuition is that the methods within the same module usually collaborate to complete certain tasks, thus tend to more frequently call each other during runtime.
Assuming that edge $\left(i, j\right)$ in the call graph $G = \left( V, E \right)$ have weights $w_{ij}$ (\ie method $i$ invoke method $j$ for $w_{i j}$ times), the quality function $Q$ of graph $G$ is as follows:
\begin{equation}
Q=\frac{1}{W} \sum_{i, j}\left[w_{i j}-\frac{k_i k_j}{W}\right] \delta\left(c_i, c_j\right)
\end{equation}
Where $W = \sum_{i,j} w_{ij}$ is the sum of all edge weights in the graph, $k_i = \sum_{j} w_{ij}$ is the weighted degree of vertex $i$, $c_i$ is the module to which vertex $i$ belongs. $\delta(c_i, c_j)$ is an indicator function whose value equals to 1 if  $c_i = c_j$, and 0 otherwise.
The value of quality $Q$ ranges between $[-1, 1]$, a higher $Q$ indicates a better partition, implying that the connections within the detected functional modules are tighter, while the connections between the modules are looser.
Based on the definition of $Q$, the Leiden algorithm is able to continuously search the optimal partition of the call graph.
We then regard each clustered call subgraph as a candidate function module.

\prompt{Prompt 1: Module-Level Knowledge Extraction}{1}{
You are an expert software analyst with deep knowledge of program structure.\\

\vspace{-0.1cm}
\textbf{\# Goal}

As a senior software engineer specializing in code analysis, generate a comprehensive functionality summary report for a method call subgraph from a program execution. This report should include an overview of the key methods in the subgraph and their calling relationships. \\

\vspace{-0.1cm}
\textbf{\# Call Graph}

Methods: <Information of the method nodes in the call graph>

Calls: <Call relationships between method nodes> \\

\vspace{-0.1cm}
\textbf{\# Report Structure}

The report should include the following sections:

\hspace{10pt}- TITLE: A short but specific title representing the main functionality of the call graph ...

\hspace{10pt}- SUMMARY: An executive summary of the subgraph's overall structure ...

\hspace{10pt}- DETAILED FINDINGS: A list of insights about the call graph ... \\

\vspace{-0.1cm}
\textbf{\# Example}

Methods:

id,className:methodName(startLine-endLine),comment

1,Main:main(1-10),The main entry point of the program

...

Calls:

id,source,target

1,Main:main(1-10),DataProcessor:processData(11-20)

...

Output:

"title": "Data Processing and Validation Flow"

"summary": "This subgraph demonstrates a simple data processing and validation flow ..."

"findings":

\hspace{10pt} "summary": "Central role of processData method"

\hspace{10pt} "explanation": "The processData method is not only responsible for data processing but also ensures data validity by calling ..."
 
\hspace{10pt}...

}

\emph{Model Size Optimization.}
Nevertheless, the clustering process of the Leiden algorithm is greedy, which may produce functional modules that are either too large or too small.
To address this issue, we have imposed a limit on the maximum size of modules that can be detected by the Leiden algorithm.
Additionally, we further designed an algorithm to prevent the formation of excessively small modules.
For each undersized module, we first select the most closely connected adjacent module as the target module, which is determined by the magnitude of the total edge weights between modules.
Subsequently, we migrate all method nodes from the undersized modules to the target module, thus eliminating 
excessively small modules without losing node information.

\emph{Knowledge Extraction.}
After module detection, we match each functional module in the dynamic call graph with the static code repository, which provides additional attributes for each method node in the graph, such as method comments, source code, \etc
Finally, we leverage the graph structure data analysis capabilities~\cite{chen2024graph,edge2024graphrag} of LLMs to extract knowledge at the functional module level, which is expressed in the form of natural language module summary as shown in Figure \ref{fig:approach}.
An example prompt for the LLM is shown in Prompt \ref{prompt:1}.
For each module (represented by a call subgraph) to be summarized, we transform the method nodes and call edges in the graph into the text format that can be interpreted by LLMs.
Acting as an \emph{expert software analyst}, the LLM is instructed to \emph{``generate a comprehensive functionality summary report''} for each functional module.
The generated report in \emph{\# Report Structure} should consists of three sections: 
\emph{TITLE} names the overall functionality of the module; 
\emph{SUMMARY} summarizes the overall structure of the module and how methods interact with each other;
and \emph{DETAILED FINDINGS} is a list of insights that elaborate on the behavior of some important methods.

\textbf{Method- \& Chunk-Level Knowledge Extraction.}
In this stage, our objective is to acquire knowledge at the method and chunk levels within the software.
We observe that existing LLM-based FL techniques typically require method-level semantic information to search for suspicious locations.
However, the semantic information maintained by these techniques for individual methods is limited to the internal scope of the method, without considering the runtime context in which the method is executed.
For instance, \autofl~\cite{kang2024autofl} directly utilizes the method code and comments as information sources, \agentfl~\cite{qin2024agentfl} further employs LLMs to enhance method comments, but the semantic information remains confined within a few methods.
To address this, we propose to consider the runtime context of the method when extracting method-level knowledge.
For each method, its context is represented by the functional module summary (we have obtained in the previous stage) where the method is located.

Nevertheless, due to the characteristics of object-oriented languages~\cite{korson1990understanding} and the prevalence of code clone~\cite{roy2009comparison}, many methods in code base may exhibit similarity in overall functionality while differing in internal logical details.
In such situations, CS based solely on method-level semantics may struggle to distinguish subtle differences between similar methods, thereby compromising the FL performance.
Consequently, we utilized LLM to summarize the detailed workflow of each method, thus decomposing the overall functionality of a method into finer-grained chunk-level units.
Here, we define a chunk as a program logic unit composed of one or more continuous statements in a method.
With the chunk-level knowledge, \toolname can differentiate similar methods through exploiting fine-grained variances in their detailed implementations, thereby further improving the FL accuracy.

An example prompt is shown in Prompt \ref{prompt:2}, for each method, we generate both method- and chunk-level knowledge in one round of conversation with the LLM.
In \emph{\# Example}, it can be observed that the semantic information for each method is derived not only from the \emph{Method Code} and \emph{Developer Comment}, but also includes the corresponding \emph{Module Context}.
As an expert \emph{code summarizer}, the LLM is tasked to \emph{``describe the key functionality and provide a detailed walkthrough of the workflow''} for each method.
The generated report in \emph{\# Report Structure} has two parts, where \emph{FUNCTIONALITY} provides a detailed description of the method's functionality and its role within the module context, and \emph{DESCRIPTION} is a list of paragraphs that explain the method's workflow in detail, which is used as the chunk-level knowledge of this method.

\prompt{Prompt 2: Method- \& Chunk-Level Knowledge Extraction}{2}{
You are a code summarizer with deep knowledge of different programming languages.\\

\vspace{-0.1cm}
\textbf{\# Goal}

As a senior software engineer specializing in code summarization, your task is to generate comprehensive documentation for a given method. This documentation should succinctly describe the key functionality of the method and provide a detailed walkthrough of its workflow.\\

\vspace{-0.1cm}
\textbf{\# Provided Information}

Method Code: <The method code to be summarized>

Developer Comment: <Any additional comments or insights provided by the developer, if available>

Module Context: <The broader context in which the method operates, if available> \\

\vspace{-0.1cm}
\textbf{\# Report Structure}

The report should include the following sections:

\hspace{10pt}- FUNCTIONALITY: A summary of the method, its role within the broader module context ...

\hspace{10pt}- DESCRIPTION: A list of detailed step-by-step explanation of the method's workflow.\\

\vspace{-0.1cm}
\textbf{\# Example}

Method Code:

\texttt{public void processTasks(List<Task> tasks) \{}

\hspace{10pt}\texttt{while (!tasks.isEmpty()) \{}
    
\hspace{20pt}\texttt{Task task = getNextTask(tasks);}
        
\hspace{20pt}\texttt{handleTask(task);}
        
\hspace{20pt}...

Developer Comment:

Handle a list of tasks by some order. Return when all tasks are done.

Module Context:

"title": "Data Processing and Validation Flow"

"summary": "This subgraph demonstrates a simple data processing and validation flow ..."

...

Output:

"functionality": "Manages the sequential processing of tasks in a list until all are completed."

"description":

\hspace{10pt} "Initially checks if the task list is not empty. If tasks remain, it retrieves the next task ..."

\hspace{10pt} "Each retrieved task is then processed by the handleTask method"
 
\hspace{10pt} ...

}

\subsection{Query Generation}

As shown in the upper left of Figure \ref{fig:approach}, the objective of the \emph{Query Generation} phase is to provide the \emph{Multi-Granularity Queries} as the input for the following \emph{Fault Retrieval} process.
Specifically, to more comprehensively depict the characteristics and context of suspicious software functionalities, \toolname generates three queries within different granularities (\ie module, method, and chunk).
However, reconsidering the example in Figure \ref{fig:motiv}, the fault information typically only contains the software’s expected behavior (Failed Test Case), actual behavior (Test Output, which is sometimes not available), and a small portion of the execution path (such as the exception stack trace), while the vast majority of the software’s internal logic remains a black box.
This prevents LLM from comprehensively understanding the root cause, let alone generating high-quality queries.
To mitigate this problem, we propose to utilize the intrinsic decision-making ability~\cite{jin2024llms} of the LLM.
Given the constructed \emph{Software Knowledge Base}, we instruct the LLM to autonomously decide whether to retrieve the relevant functional module knowledge until it generates queries with abundant confidence.
The LLM prompt is shown in Prompt \ref{prompt:3}.

For each failed test, we instruct the LLM to \emph{``either identify the potential faulty functionality or
request additional information as needed''}.
As described in \emph{\# Response Format}, if the LLM lacks sufficient information to make a determination, it is required to respond with a \emph{request} which describes the needed functional module details.
In this condition, we use semantic search to find the most relevant module summary from the module index and add it to the \emph{Module Details} section within \emph{\# Fault Information}.
Otherwise, the LLM is asked to generate three fields \emph{module}, \emph{method}, and \emph{chunk} to depict the suspicious functionality in the software at different levels of granularity.
These fields will be used as queries for the subsequent multi-granularity code search process.

\subsection{Fault Retrieval}
\label{subsec:retrieval}
As illustrated in the upper right of Figure \ref{fig:approach}, the \emph{Fault Retrieval} phase takes the \emph{Multi-Granularity Queries} as input, which aims to pinpoint the suspicious buggy methods from the perspective of semantic code search (CS).
To achieve this, we leverage the text embedding~\cite{incitti2023embedding} technique which is widely used in CS~\cite{gu2018deep}.
Text embedding models map natural/program language text into a high-dimensional vector space, where semantically similar texts are positioned closer to each other.
For the fault localization task, our insight is that the suspicious functionalities derived from LLM analysis and the actual buggy location can be semantically similar.
Specifically, the workflow of \emph{Fault Retrieval} consists of two steps: multi-granularity retrieval and suspicious method voting, we will introduce them in the following sections.

\prompt{Prompt 3: Query Generation}{3}{
You are a Software Diagnostics Specialist specializing in software fault analysis and fault localization.\\

\vspace{-0.1cm}
\textbf{\# Goal}

Your task is to analyze the provided fault information and either identify the potentially faulty functionality or request additional information as needed.\\

\vspace{-0.1cm}
\textbf{\# Fault Information}

Test Case Code: <Source Code of the Failed Test Case>

Exception Stack Trace: <Exception Stack Trace>

Test Output: <Test Output>

Module Details: <The Retrieved Module Summaries>\\

\vspace{-0.1cm}
\textbf{\# Analysis Process}

1. Analyze provided information and interactions between methods that might cause the fault.

2. Determine if sufficient information is available to pinpoint the likely cause of the fault.

3. If information is insufficient, formulate a specific question to gather necessary details.

4. If information is sufficient, describe the potentially faulty functionality at various levels of detail.\\

\vspace{-0.1cm}
\textbf{\# Response Format}

\hspace{10pt}- If you have sufficient information to identify the potentially faulty functionality, the response should contain `module', `method', and `chunk' fields, describing the likely location and nature of the bug at increasing levels of detail:

\hspace{20pt}"module": "<A description of the likely functional module where the bug resides>"

\hspace{20pt}"method": "<A description of the functionality in the software which may cause the fault>"
  
\hspace{20pt}"chunk": "<A more detailed description of the specific code logic that is likely causing the bug>"

\hspace{10pt}- If you lack sufficient information to make a determination, the response should contain a `request' field, specifying the additional information needed to complete the analysis:

\hspace{20pt}"request": "<Module details or specific information you needed to better understand the fault>"

}

\textbf{Multi-Granularity Retrieval.}
We first build embedding indexes for program elements at different levels using a text embedding model, where the embedded textual content is derived from the knowledge acquired in the previous Section \ref{subsec:knowledge}.
Formally, we represent all program elements as $E=\{G, M, S\}$, where $G=\{g_{1}, \dots, g_{|G|}\}$ denotes function modules (\ie dynamic call subgraphs), $M=\{m_1, \dots, m_{|M|}\}$ represents methods, and $S=\{S^{m_{1}}, \dots, S^{m_{|M|}}\}$ comprises all chunks for all methods, with $S^{m_{i}} = \{s^{ m_{i}}_{1}, \dots, s^{ m_{i}}_{|S^{m_{i}}|}\}$ being the chunks in method $m_i$.
We then utilize a text embedding model $\sigma$ to vectorize the knowledge text, which results in the \emph{Module Index} $I^{G} = \{ \sigma(\kappa(g_{1})), \dots, \sigma(\kappa(g_{|G|}) \}$, \emph{Method Index} 
$I^{M} = \{ \sigma(\kappa(m_{1})), \dots, \sigma(\kappa(m_{|M|})) \}$ , and \emph{Chunk Index} $I^{S}= \{\sigma(\kappa(S^{m_{1}})), \dots, \sigma(\kappa(S^{m_{|M|}}))\}$ as shown in Figure \ref{fig:approach}.
Here, $\kappa$ maps each program element to its corresponding knowledge summary extracted by the LLM.

Next, let $U = \{u_{i} = (c_{i}, f_{i}, l_{i}) | i = 1, 2, \dots, N \}$ denote the multi-granularity queries generated in the \emph{Query Generation} phase, where $c$, $f$, and $l$ are queries respectively corresponding to module-, method-, and chunk-level suspicious functionalities, $N$ is the number of failed test cases.
The knowledge base actually maintains three granularities of knowledge $k_{m} = ( \kappa ( \phi (m)), \kappa (m), \kappa (\varphi(m)))$ for each method $m$, where $\phi : M \mapsto G$ maps a method to the module in which it is located, and $\varphi : M \mapsto S$ maps each method $m$ to the set of chunks $S^{m_{i}}$ within that method.
Given the above condition, the retrieval process can be formally described as:

\begin{align}
    & \hat{M}_{i} = \{ \hat{m}_{1}, \dots, \hat{m}_{|\hat{M}_{i}|}\} = \text{Retrieve}(I^{M}, f_{i}, \lambda) \label{eq:2}\\
    & \hat{G}_{i} = \{ \hat{g}_{1}, \dots, \hat{g}_{|\hat{G}_{i}|}\} = \text{Retrieve}(I^{G}_{p}, c_{i}, \lambda), \quad I^{G}_{p} = \{ \phi(\hat{m}_{1}), \dots, \phi(\hat{m}_{|\hat{M}_{i}|}) \} \label{eq:3}\\
    & \hat{S}_{i} = \{ \hat{s}_{1}, \dots, \hat{s}_{|\hat{S}_{i}|}\} = \text{Retrieve}(I^{S}_{p}, l_{i}, \lambda), \quad I^{S}_{p} = \varphi(\hat{m}_{1}) \cup \dots \cup \varphi(\hat{m}_{|\hat{M}_{i}|}) \label{eq:4}\\
    & \hat{E} = \{ (\hat{G}_{i}, \hat{M}_{i}, \hat{S}_{i} ) | i = 1, \dots, N\} \label{eq:5}
\end{align}

The Retrieve function measures the semantic similarity between the query and program element summaries using cosine similarity~\cite{thongtan2019sentiment}, and returns the suspicious program elements in descending order of similarity.
Here, $\lambda$ is a hyper-parameter used to control the number of retrieved program elements.
Specifically, for each failed test case, Formula \ref{eq:2} is used to retrieve suspicious methods, while Formulas \ref{eq:3} and \ref{eq:4} are used to retrieve suspicious modules and chunks, respectively.
Formula \ref{eq:5} denotes the results from all failed test cases.
It is noteworthy that we use the results from method retrieval $\hat{M}_{i}$ to narrow down the size of module and chunk index ($I^{G}_{p}$ and $I^{S}_{p}$), which can improve retrieval efficiency by excluding a large number of irrelevant program elements.

\textbf{Suspicious Method Voting.}
After obtaining all retrieved program elements $\hat{E}$ in formula \ref{eq:5}, we are supposed to aggregate the retrieval results from multiple queries and granularities, which finally results in a list of methods sorted in descending order of suspiciousness.
Inspired by the Borda count~\cite{emerson2013original} in social science, we designed a voting mechanism to elect the more suspicious methods.
Our insight is that if a method (including its related module and chunks) is semantically related to more queries, then this method is more likely to be the actual fault location.
Formally, for the set of all retrieved methods $M^{*} = \hat{M}_{1} \cup \dots \cup \hat{M}_{N}$, the suspicious score of an individual method $m^{*} \in M^{*}$ is calculated using the following formula:
\begin{align}
    \text{Score}(m^{*}) &= \sum_{i=1}^{N}\sum_{j=1}^{|\hat{G}_{i}|}\omega(\hat{g}_{ij}) + \sum_{i=1}^{N}\sum_{j=1}^{|\hat{M}_{i}|}\omega(\hat{m}_{ij}) + \sum_{i=1}^{N}\sum_{j=1}^{|\hat{S}_{i}|}\omega(\hat{s}_{ij}) \label{eq:6}\\
    \omega(\hat{g}_{ij}) &=
    \begin{cases}
        \epsilon(\hat{g}_{ij}), & \text{method } m^{*} \text{ belongs to module } \hat{g}_{ij} \\ 
        0, & \text{otherwise}
    \end{cases} \label{eq:7} \\
    \omega(\hat{m}_{ij}) &=
    \begin{cases}
        \epsilon(\hat{m}_{ij}), & m^{*} = \hat{m}_{ij} \\ 
        0, & \text{otherwise}
    \end{cases} \label{eq:8} \\
    \omega(\hat{s}_{ij}) &=
    \begin{cases}
        \epsilon(\hat{s}_{ij}), & \text{chunk } \hat{s}_{ij} \text{ belongs to method } m^{*} \\ 
        0, & \text{otherwise}
    \end{cases} \label{eq:9}
\end{align}
$\epsilon$ is computed during the retrieval process with a value range of $[-1, 1]$, representing the semantic similarity between a program element and its corresponding query.
As shown in Formula \ref{eq:6}, the suspiciousness of a method $m^{*}$ not only stems from its similarity to method-level queries (calculated by the second term of the formula), but also from the degree to which the associated module and chunks of the method match to the corresponding queries (the first and third terms).
After obtaining the suspiciousness score for each method $m^{*}$, all methods in $M^{*}$ will be ranked in descending order.
Finally, we follow previous studies~\cite{kochhar2016practitioner,kang2024autofl} to generate user-friendly explanations for the localization results through the LLM.

\section{Experiment}

\subsection{Research Question}
To comprehensively evaluate the performance of \toolname, we design three research questions:

\textbf{RQ1: The effectiveness of \toolname.}
How effective is \toolname in method-level fault localization compared to other state-of-the-art code navigation-based FL techniques?

\textbf{RQ2: The impact of different components in \toolname.}
We conduct an ablation study to investigate how different components contribute to the overall performance of \toolname.

\textbf{RQ3: Effects of different hyper-parameters on \toolname.}
 We explore the sensitivity of \toolname to different parameter settings.


\subsection{Benchmark}

\begin{wraptable}[25]{r}{0.6\textwidth}
\centering
\caption{Benchmark Information.}
\label{tab:benchmark}
\resizebox{0.6\textwidth}{!}{
\begin{tabularx}{0.8\textwidth}{p{0.4cm}|p{2.4cm}p{3.5cm}|>{\raggedleft\arraybackslash}X>{\raggedleft\arraybackslash}X>{\raggedleft\arraybackslash}X}
\midrule
& \textbf{Project} & \textbf{Name}          & \textbf{\#Bug} & \textbf{LoC} & \textbf{NoM} \\
\midrule
\multirow{6}{*}{\begin{sideways}Defects4J V1.2.0\end{sideways}} & Chart            & jfreechart             & 25             & 207K         & 14,925        \\
& Closure          & closure-compiler       & 125            & 123K         & 18,052        \\
& Lang             & commons-lang           & 61             & 56K          & 8,325         \\
& Math             & commons-math           & 99             & 164K         & 20,503        \\
& Mockito          & mockito                & 35             & 20K          & 2,522         \\
& Time             & joda-time              & 25             & 61K          & 3,993         \\
\midrule
\multirow{12}{*}{\begin{sideways}Defects4J V2.0.0\end{sideways}} & Closure          & closure-compiler       & 41             & 123K         & 18,052        \\
& Cli              & commons-cli            & 37             & 4K           & 1,400         \\
& Codec            & commons-codec          & 15             & 5K           & 1,655         \\
& Collections      & commons-collections    & 3              & 61K          & 5,004         \\
& Compress         & commons-compress       & 44             & 30K          & 5,666         \\
& Csv              & commons-csv            & 15             & 2K           & 592          \\
& Gson             & gson                   & 14             & 10K          & 1,271         \\
& JacksonCore      & jackson-core           & 19             & 24K          & 3,889         \\
& JacksonDatabind  & jackson-databind       & 39             & 65K          & 13,861        \\
& JacksonXml       & jackson-dataformat-xml & 6              & 6K           & 523          \\
& Jsoup            & jsoup                  & 88             & 4K           & 2,964         \\
\midrule
\multirow{7}{*}{\begin{sideways}GrowingBugs\end{sideways}} & IO               & commons-io             & 22             & 11K          & 3,646         \\
& Validator        & commons-validator      & 11             & 16K          & 1,292         \\
& Javapoet         & javapoet               & 16             & 1K           & 139          \\
& Zip4j            & zip4j                  & 47             & 7K           & 1,442         \\
& Spoon            & spoon                  & 16             & 35K          & 6,067         \\
& Markedj          & markedj                & 13             & 2K           & 168          \\
& Dagger\_core     & dagger-core            & 19             & 3K           & 382          \\
\midrule
& \multicolumn{2}{c|}{\textbf{Overall}} & \textbf{835} & \textbf{1,066K} & \textbf{139K} \\
\midrule
\end{tabularx}
}
\end{wraptable}
In order to comprehensively evaluate the effect of \toolname on different software projects, we utilize the widely used software defect benchmark Defects4J~\cite{just2014defects4j}.
We reproduce 691 bugs from 16 open-source Java projects, of which Defects4J V1.2.0 contains 370 bugs from 6 projects, and Defects4J V2.0.0 adds 321 bugs from 11 projects.
To further enrich the types of software projects, we collected extra 144 reproducible bugs of 7 projects from the GrowingBugs~\cite{jiang2022growingbugs} benchmark.


Table \ref{tab:benchmark} shows the detailed information of the benchmark.
The columns ``Project'' and ``Name'' represent the abbreviation and full name of the software project, respectively.
The column ``\#Bug'' represents the number of bugs in the project.
``LoC'' and ``NoM'' show the total number of lines of code and the number of methods of the software project, respectively, where the number of methods does not include methods without bodies such as abstract methods.
Finally, the benchmark we used to conduct the experiments comprises 835 bugs from 23 Java projects.

\subsection{Baseline}
We compare \toolname with two state-of-the-art fault localization approaches, \autofl~\cite{kang2024autofl} and \agentfl~\cite{qin2024agentfl}.
Both of them focus on localizing the fault location from the entire software project through code navigation.

\autofl~\cite{kang2024autofl} utilizes the function calling capabilities of LLMs for fault localization.
Given functions such as \texttt{get\_class\_covered}, \texttt{get\_method\_covered}, and \texttt{get\_code\_snippet}, the LLM is required to autonomously decide whether to invoke functions to acquire additional information in each round of dialogue, and gradually navigating to the buggy location.
It is worth noting that \autofl employs repeated runs to reduce the perplexity of results. For the sake of fairness, we set the number of runs for all FL tools to 1 in this paper.

\agentfl~\cite{qin2024agentfl} defines the fault localization task as a standard operating procedure, which is decomposed into three steps: fault comprehension, codebase navigation, and fault confirmation.
In each step, \agentfl employs agents capable of utilizing various tools to address specific subtasks.
\agentfl incorporates a document-guided search strategy for codebase navigation, which utilizes both existing documents and LLM-enhanced documents to identify classes and methods related to the potential root causes.

\subsection{Metrics}
Following existing FL approaches~\cite{lou2021grace}, we use three metrics to evaluate the method-level FL performance in this paper:

\textbf{Top-N Recall (Top-N).}
For a single bug version $b \in B$, Top-N indicates whether there is a buggy method $m \in M$ in the first $N$ methods of the method ranking list $L$ provided by the fault localization approach. For an entire project, Top-N is equal to the sum of the results of all bug versions. The formula of Top-N is $\text{Top-N} = \sum_{i=1}^{|B|} I_{i}$, where $I_{i}$ equals to 1 if $\exists m \in \{L_1, L_2, \dots, L_N\}, m \in M$, or 0 otherwise. In this work, we adopt $N \in \{1,5,10\}$. The higher the Top-N recall, the better the fault localization performance.

\textbf{Mean First Rank (MFR).}
MFR can be used to indicate the ranking of the first buggy method in the method ranking list. For a single buggy version, MFR is the ranking of the first buggy method in the list. For all buggy versions in the entire project, MFR is equal to the mean of the results of all buggy versions. The formula of MFR is $\text{MFR} = \frac{1}{|B|} \sum_{i=1}^{|B|} rank_{i}$. If no buggy method is found within the recall set of size $N$, the rank value is assigned to $N+1$. The lower the MFR, the better the fault localization performance.

\textbf{Mean Average Rank (MAR).}
MAR is utilized to indicate the ranking of all buggy methods in the method ranking list. For a single buggy version, MFR is the average ranking of all buggy methods in the list. For all buggy versions in the entire project, MFR is equal to the mean of the results of all buggy versions. The formula of MAR is $\text{MAR} = \frac{1}{|B||M|} \sum_{i=1}^{|B|} \sum_{j=1}^{|M|} rank_{ij}$. The lower the MAR, the better the fault localization performance.

    

\subsection{Implementation}
We implement \toolname based on  LlamaIndex~\cite{llamaindex}.
We use ChromaDB~\cite{chromadb} as the vector store and tree-sitter~\cite{treesitter} for AST-based parsing and program element extraction.
For inspecting software dynamic behavior, we conduct program instrumentation through a self-implement Java agent~\cite{instrument} and build the dynamic method call graph on the fly with JGraphT~\cite{jgrapht}.

We set the following alterable parameters for the default version of \toolname.
In \emph{Module-Level Knowledge Extraction} of Section \ref{subsec:knowledge}, the maximum and minimum module sizes are set to 15 and 5 respectively.
The number of retrieved elements in \emph{Fault Retrieval} of Section \ref{subsec:retrieval} is set to $\lambda = 50$.
We use DeepSeek-V2.5~\cite{deepseekv2.5} as the LLM backend, which has outstanding performance in code ability while also provides a preferential price (7 million input tokens or 3.5 million output tokens for 1\$), the temperature is set to 1.0.
For the same reason, we choose jina-embeddings-v2-base-en~\cite{jina} as the default text embedding model.
\section{Evaluation}
\label{sec:eval}

\subsection{RQ1: Effectiveness of \toolname}

In Table \ref{tab:rq1}, \toolname is compared with other LLM-based techniques on method-level FL effectiveness.
Overall, \toolname achieves the best performance, which localizes 324, 532, and 566 bugs in the top 1, 5, and 10 positions, respectively.
The MFR is 14.15 and the MAR is 14.97, indicating that \toolname can rank the buggy methods at relatively high positions.
The effect of \agentfl is inferior to \toolname, which localizes 256, 384, and 393 methods in the top 1, 5, and 10 positions, with an MFR of 26.80 and MAR of 26.90.
\autofl performed the worst among the three approaches, with Top1, Top3, and Top5 scores of 206, 253, and 254, respectively, and MFR of 35.78 and 38.48.
\toolname shows 57.3\% and 26.6\% increases in Top1 compared to \autofl and \agentfl, respectively.
For the Top5 accuracy, the increases are even more significant, reaching 110.3\% and 38.5\%.
Such results are consistent with our expectation that the performance of FL techniques is proportional to the amount of software information they can exploit.
Specifically, \autofl only utilized the most basic information such as the names of the covered classes and methods;
\agentfl further incorporates documentation generated by developers or LLMs, providing more detailed knowledge about software functionality;
\toolname provides the richest information through a software knowledge base, which supplies runtime information about bugs, offering a basis for more accurate fault localization.

\begin{table}[t]
\caption{Results of \toolname and other code navigation based FL approaches.}
\label{tab:rq1}
\resizebox{\linewidth}{!}{
\begin{tabularx}{1.2\linewidth}{p{19mm}>{\raggedleft\arraybackslash}p{7mm}|>{\raggedleft\arraybackslash}p{4mm}>{\raggedleft\arraybackslash}p{4mm}>{\raggedleft\arraybackslash}p{4mm}>{\raggedleft\arraybackslash}X>{\raggedleft\arraybackslash}X|>{\raggedleft\arraybackslash}p{4mm}>{\raggedleft\arraybackslash}p{4mm}>{\raggedleft\arraybackslash}p{4mm}>{\raggedleft\arraybackslash}X>{\raggedleft\arraybackslash}X|>{\raggedleft\arraybackslash}p{4mm}>{\raggedleft\arraybackslash}p{4mm}>{\raggedleft\arraybackslash}p{4mm}>{\raggedleft\arraybackslash}X>{\raggedleft\arraybackslash}X}
\hline
\multirow{2}{*}{\textbf{Project}} & \multirow{2}{*}{\textbf{Bugs}} & \multicolumn{5}{c|}{\textbf{\autofl}}                                       & \multicolumn{5}{c|}{\textbf{\agentfl}}                                     & \multicolumn{5}{c}{\bftoolname}                                           \\ \cline{3-17} 
                                  &                                & \textbf{T1}  & \textbf{T5}  & \textbf{T10} & \textbf{MFR} & \textbf{MAR} & \textbf{T1}  & \textbf{T5}  & \textbf{T10} & \textbf{MFR} & \textbf{MAR} & \textbf{T1}  & \textbf{T5}  & \textbf{T10} & \textbf{MFR}   & \textbf{MAR}   \\ \hline
\rowcolor{tableblue} Chart           & 25  & 6   & 9   & 9   & 33.16 & 34.16 & 12  & 18  & 18  & 15.40 & 15.74 & 15  & 23  & 23  & 4.24  & 5.04   \\
Closure         & 166 & 3   & 4   & 4   & 49.79 & 50.39 & 27  & 53  & 61  & 24.17 & 24.50 & 35  & 61  & 73  & 29.34 & 30.25  \\
\rowcolor{tableblue} Lang            & 61  & 27  & 31  & 31  & 25.67 & 28.27 & 38  & 46  & 46  & 13.49 & 13.61 & 36  & 52  & 55  & 5.20  & 5.42   \\
Math            & 99  & 43  & 56  & 56  & 22.90 & 26.04 & 48  & 64  & 64  & 18.96 & 19.11 & 46  & 74  & 75  & 8.80  & 9.53   \\
\rowcolor{tableblue} Mockito         & 35  & 15  & 16  & 16  & 28.17 & 38.89 & 16  & 18  & 18  & 25.37 & 25.59 & 15  & 23  & 24  & 13.60 & 14.74  \\
Time            & 25  & 10  & 11  & 11  & 29.08 & 34.25 & 12  & 14  & 14  & 23.08 & 23.10 & 7   & 13  & 13  & 19.16 & 20.14  \\
\rowcolor{tableblue} Cli             & 37  & 6   & 10  & 10  & 37.25 & 37.25 & 12  & 22  & 22  & 21.78 & 21.96 & 15  & 26  & 30  & 6.24  & 7.82   \\
Codec           & 15  & 4   & 5   & 5   & 33.21 & 33.60 & 5   & 12  & 12  & 11.80 & 11.93 & 6   & 13  & 14  & 3.00  & 3.84   \\
\rowcolor{tableblue} Collections     & 3   & 0   & 0   & 0   & 50.33 & 50.33 & 0   & 0   & 0   & 51.00 & 51.00 & 1   & 1   & 1   & 34.33 & 34.33  \\
Compress        & 44  & 16  & 20  & 20  & 28.45 & 30.71 & 17  & 28  & 28  & 19.55 & 19.55 & 18  & 30  & 33  & 9.16  & 10.52  \\
\rowcolor{tableblue} Csv             & 15  & 4   & 4   & 4   & 37.67 & 41.00 & 4   & 5   & 5   & 34.47 & 34.47 & 9   & 11  & 11  & 5.07  & 5.37   \\
Gson            & 14  & 4   & 6   & 6   & 29.79 & 33.87 & 6   & 7   & 7   & 26.07 & 26.14 & 8   & 11  & 11  & 6.64  & 6.75   \\
\rowcolor{tableblue} JacksonCore     & 19  & 3   & 6   & 6   & 35.47 & 38.00 & 7   & 11  & 11  & 22.42 & 22.73 & 9   & 15  & 15  & 8.05  & 8.75   \\
JacksonDatabind & 39  & 4   & 5   & 5   & 44.54 & 50.08 & 4   & 6   & 7   & 42.31 & 42.32 & 5   & 15  & 16  & 28.13 & 28.55  \\
\rowcolor{tableblue} JacksonXml      & 6   & 1   & 1   & 1   & 42.50 & 42.50 & 1   & 1   & 1   & 42.67 & 42.67 & 3   & 4   & 4   & 16.50 & 16.50  \\
Jsoup           & 88  & 25  & 25  & 25  & 36.75 & 40.57 & 21  & 37  & 37  & 30.31 & 30.34 & 31  & 52  & 56  & 16.83 & 17.78  \\
\rowcolor{tableblue} IO              & 22  & 12  & 18  & 18  & 10.36 & 14.15 & 0   & 0   & 0   & 51.00 & 51.00 & 13  & 21  & 21  & 2.18  & 2.30   \\
Validator       & 11  & 2   & 3   & 3   & 34.33 & 34.33 & 5   & 9   & 9   & 10.45 & 10.45 & 8   & 10  & 10  & 5.91  & 6.16   \\
\rowcolor{tableblue} Javapoet        & 16  & 7   & 7   & 7   & 29.13 & 30.69 & 10  & 14  & 14  & 7.63  & 7.63  & 12  & 16  & 16  & 1.25  & 1.66   \\
Zip4j           & 47  & 10  & 10  & 10  & 40.36 & 42.28 & 6   & 8   & 8   & 42.62 & 42.65 & 15  & 29  & 30  & 14.55 & 16.09  \\
\rowcolor{tableblue} Spoon           & 16  & 3   & 3   & 4   & 38.81 & 40.22 & 2   & 7   & 7   & 29.56 & 29.56 & 6   & 8   & 9   & 19.88 & 20.63  \\
Markedj         & 13  & 0   & 0   & 0   & 51.00 & 51.00 & 1   & 2   & 2   & 43.38 & 43.38 & 5   & 11  & 12  & 3.31  & 6.28   \\
\rowcolor{tableblue} Dagger\_core    & 19  & 1   & 3   & 3   & 43.11 & 47.40 & 2   & 2   & 2   & 45.74 & 45.74 & 6   & 13  & 14  & 12.79 & 15.16  \\ \hline
\textbf{Total}                    & \textbf{835}                   & \textbf{206} & \textbf{253} & \textbf{254} & \textbf{35.78}    & \textbf{38.48}    & \textbf{256} & \textbf{384} & \textbf{393} & \textbf{26.80}    & \textbf{26.90}    & \textbf{324} & \textbf{532} & \textbf{566} & \textbf{14.15} & \textbf{14.97} \\
\hline
\end{tabularx}
}
\end{table}

\begin{wrapfigure}[11]{r}{0.30\textwidth}
  \centering
  \includegraphics[width=0.25\textwidth]{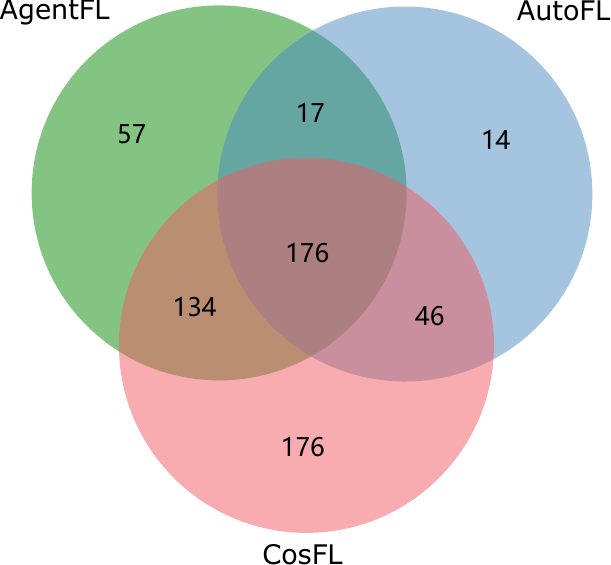}
  \caption{Overlap of Top5 results.}
  \label{fig:rq1}
  \Description{}
\end{wrapfigure}

In Figure \ref{fig:rq1}, we present the overlap of Top5 results among different FL approaches.
Particularly, we employed the Top5 metric to measure the comprehensive localization performance, as a previous study~\cite{kochhar2016practitioner} indicated that over 70\% of practitioners consider localization results that rank buggy elements within the top five positions to be successful.
As shown in the figure, we observe that the results of \toolname substantially cover the other two, identifying 80.7\% (310 out of 384) of the bugs localized by \agentfl and 87.7\% (222 out of 253) of \autofl.
Moreover, \toolname independently finds 176 bugs, which illustrates the effectiveness of boosting FL through semantic code search.


\notez{
{\bf Answer to RQ1:}
\toolname shows superior performance in fault localization by ranking 324 (532) out of 835 bugs within the top 1 (5) positions, significantly outperforming other LLM-based methods.
}


\subsection{RQ2: Ablation Study}
\label{subsec:ablation}

To evaluate the contribution of different components in \toolname, we devised three variants:
1) \textbf{w/o Module Context} removes the module-level knowledge extraction process, thereby impacting both the \emph{Query Generation} and \emph{Fault Retrieval} phases.
2) \textbf{w/o Module Retrieval} eliminates module-level retrieval in \emph{Fault Retrieval}.
3) \textbf{w/o Chunk Retrieval} excludes chunk-level retrieval in \emph{Fault Retrieval}.
The effectiveness of each variant is presented in Table \ref{tab:rq2}.

In the first row of the table, we observe that module-level knowledge has a significant impact on \toolname.
Interestingly, after removing the module context, the Top1 value drops sharply by 80, while the Top5 remains relatively unchanged and the Top10 even increases slightly by 32.
This phenomenon reflects the influence of module-level knowledge from both sides:
On the one hand, additional information reduces the confusion of the LLM, leading to more precise descriptions of faulty functionalities, thereby positioning the buggy methods higher in the list.
On the other hand, the reduction in ambiguity somewhat narrows the search scope, decreasing the likelihood of a small fraction of bugs being localized.
From rows 2 to 3, we can see that both module- and chunk-granularity retrieval enhance the effectiveness of \toolname, contributing respectively to Top1 improvements of 12 and 19, and Top10 improvements of 9 and 23.
Notably, the chunk-level retrieval plays a more critical role, offering an MRR contribution of 1.02, in contrast to 0.40 from module retrieval.
This indicates that subtle variations in the detailed chunks between different methods can serve as a valid basis for determining the fault location.

\begin{table}[t]
\caption{Ablation study results.}
\label{tab:rq2}
\resizebox{0.8\textwidth}{!}{
\begin{tabular}{c|c|l|ccccc}
\hline
\multicolumn{1}{l|}{\textbf{Project}} & \multicolumn{1}{l|}{\textbf{Bugs}} & \multicolumn{1}{l|}{\textbf{Variants}} & \textbf{Top1} & \textbf{Top5} & \textbf{Top10} & \textbf{MFR} & \textbf{MAR} \\ \hline
\multirow{4}{*}{Overall}              & \multirow{4}{*}{835}               & w/o Module Context & 244 \colorbox{tbred}{$\downarrow 80$} & 531 \colorbox{tbred}{$\downarrow \text{\phantom{0}}1$} & 598 \colorbox{tbgreen}{$\uparrow 32$} & 14.38 \colorbox{tbred}{$\uparrow 0.23$} & 15.15 \colorbox{tbred}{$\uparrow 0.18$} \\
 & & w/o Module Retrieval  & 312 \colorbox{tbred}{$\downarrow 12$} & 519 \colorbox{tbred}{$\downarrow 13$} & 557 \colorbox{tbred}{$\downarrow \text{\phantom{0}}9$} & 14.68 \colorbox{tbred}{$\uparrow 0.53$} & 15.37 \colorbox{tbred}{$\uparrow 0.40$} \\
 & & w/o Chunk Retrieval  & 305 \colorbox{tbred}{$\downarrow 19$} & 509 \colorbox{tbred}{$\downarrow 23$} & 543 \colorbox{tbred}{$\downarrow 23$} & 15.15 \colorbox{tbred}{$\uparrow 1.00$} & 15.99 \colorbox{tbred}{$\uparrow 1.02$} \\
 & & \cellcolor{tableblue}  Default \toolname               & \cellcolor{tableblue} 324         & \cellcolor{tableblue} 532         & \cellcolor{tableblue} 566          & \cellcolor{tableblue} 14.15        & \cellcolor{tableblue} 14.97 \\ \hline
\end{tabular}
}
\end{table}

\notez{
{\bf Answer to RQ2:}
All components have contributed to the fault localization performance of \toolname, among which the module-level knowledge has exerted a relatively more significant impact.
}

\begin{figure}[!t]
  \centering
  \includegraphics[width=0.8\linewidth]{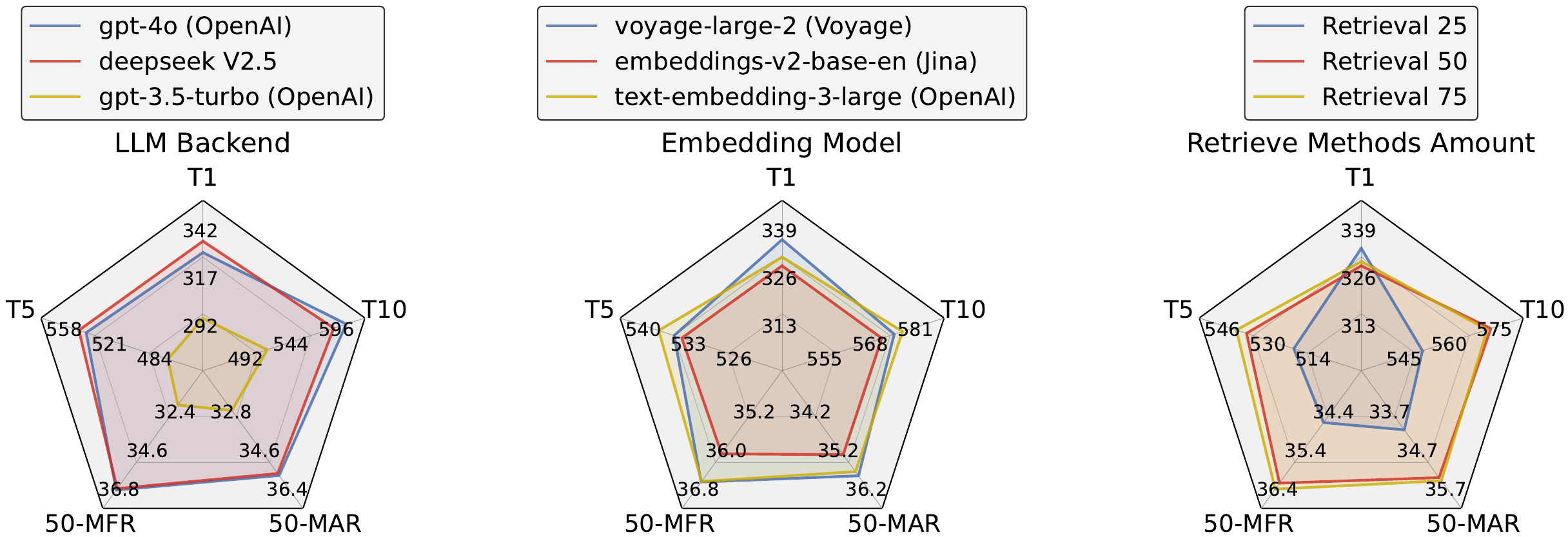}
  \caption{The impact of different hyper-parameters on \toolname.}
  \label{fig:rq3}
  \Description{rq3.}
\end{figure}

\subsection{RQ3: Sensitivity Analysis}
\label{subsec:rq3}

Figure \ref{fig:rq3} illustrates the performance of \toolname under different hyper-parameter settings, where the red line represents the default settings. 
Regarding the \textbf{LLM Backend}, we observe that LLMs with varying capabilities significantly impact \toolname's performance. For instance, \toolname (gpt-3.5-turbo) ranked 290 and 471 bugs in the Top1 and Top5, respectively. However, with gpt-4o, the Top1 and Top5 dramatically increased by 29 and 56.
Nevertheless, for LLMs with similar capacities, such as deepseek V2.5 and gpt-4o, the performance variation of \toolname is negligible.
Concerning the \textbf{Embedding Model}, we notice that \toolname's performance remained stable across different series of embedding models (i.e., Jina~\cite{jina}, Voyage~\cite{voyageai}, and OpenAI~\cite{openai}).
The performance differences introduced by \emph{text-embedding-3-large} and \emph{voyage-large-2} did not exceed 2\% across all metrics.
For \textbf{Retrieve Methods Amount}, the performance of \toolname improves as the retrieved methods amount $\lambda$ increases from 25 to 50, with gains of 14 and 19 in Top5 and Top10, respectively.
However, as $\lambda$ further increases from 50 to 75, the performance nearly plateaued. This indicates that setting $\lambda$ at a moderate level is enough to fully harness \toolname's capabilities.

\notez{
{\bf Answer to RQ3:}
\toolname shows stable performance under backend models of the same capability levels. By setting appropriate retrieve methods amount, the potential of \toolname can be further unleashed.
}

\section{Discussion}
\label{sec:dis}

\subsection{Case Study}
\label{subsec:case}
To help understand how the workflow of \toolname differs from other FL approaches, we further demonstrate two specific cases.
In the first case, we compare \toolname with \agentfl~\cite{qin2024agentfl} to demonstrate the \toolname's ability to inspect the internal logic of the software by requesting module-level knowledge.
In the second case, the comparison between \toolname and \autofl~\cite{kang2024autofl} indicates the superiority of semantic code search in expanding search scope.

\begin{figure}[!t]
  \centering
  \includegraphics[width=\linewidth]{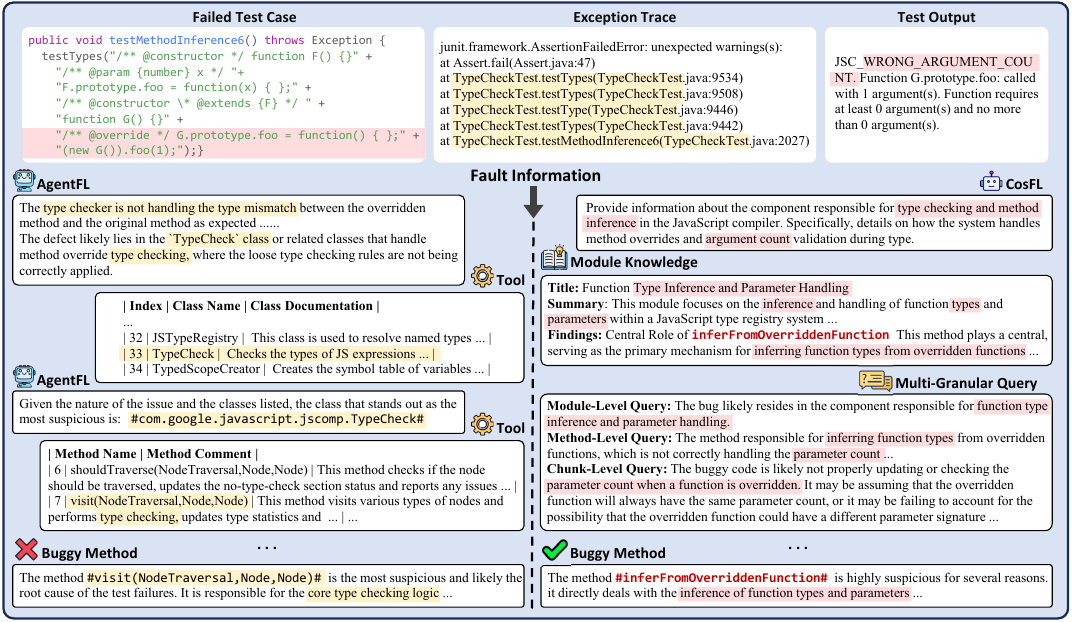}
  \caption{Comparison between \agentfl and \toolname in localizing bug Closure-41. The keywords that \agentfl and \toolname focus on are marked in  \colorbox{rqyellow}{gold} and \colorbox{rqred}{red}, respectively.}
  \label{fig:rq4-1}
  \Description{rq4-1.}
\end{figure}

\textbf{Case 1: Better Fault Understanding.}
In Figure \ref{fig:rq4-1}, we illustrate the different workflow of \agentfl and \toolname in localizing the bug Closure-41.\footnote{\url{https://storage.googleapis.com/google-code-archive/v2/code.google.com/closure-compiler/issues/issue-368.json}}
The cause of this bug is that the software throws an unexpected exception when performing a type check on the overridden function \emph{G.prototype.foo}.
To fix this issue, the developers modified the \emph{inferFromOverriddenFunction} method within the \emph{FunctionTypeBuilder} class, adding logic to enhance the parameter checking of the function.
In this instance, since \agentfl can only start from the provided \emph{Fault Information}, it is restricted to focusing on the \emph{TypeCheckTest} class which repeatedly appears in the \emph{Exception Trace}, leading to the erroneous inference that \emph{``The defect likely lies in the TypeCheck class.''}
This inference governs the follow-up localization process, misleading \agentfl to identify an unexpected method as buggy.
Conversely, to gain a deeper understanding of the root cause, \toolname starts with asking \emph{``provide information about the component responsible for type checking and method inference''} to obtain relevant module knowledge.
The returned \emph{Module Knowledge} highlights that the method \emph{inferFromOverriddenFunction} \emph{``serves as the primary mechanism for inferring function types from overridden functions''}. This discovery guides \toolname to the accurate fault location.


\begin{figure}[!t]
  \centering
  \includegraphics[width=\linewidth]{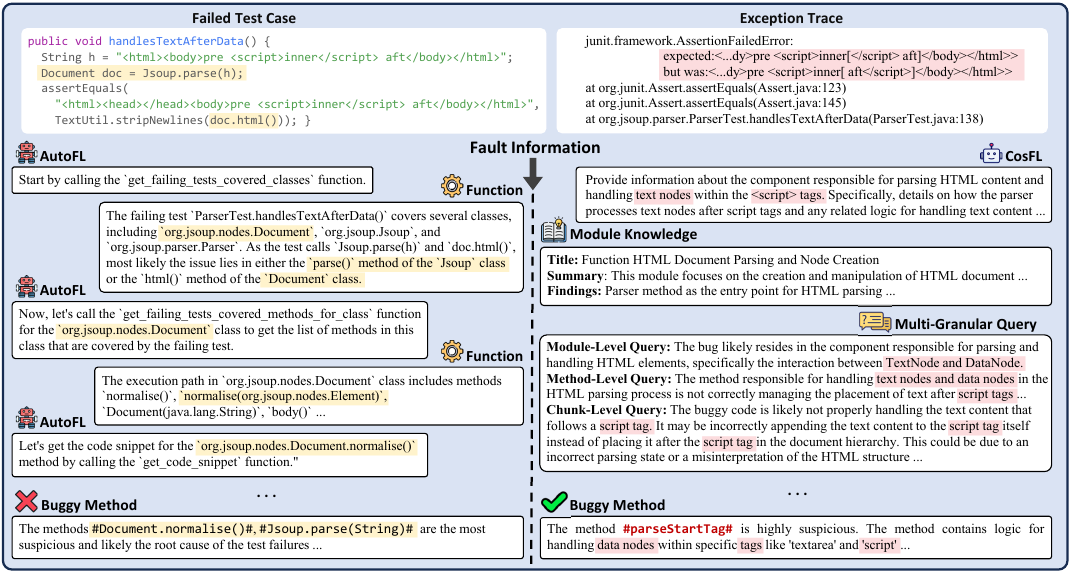}
  \caption{Comparison between \autofl and \toolname in localizing bug Jsoup-2. The keywords that \autofl and \toolname focus on are marked in  \colorbox{rqyellow}{gold} and \colorbox{rqred}{red}, respectively.}
  \label{fig:rq4-2}
  \Description{rq4-2.}
\end{figure}

\textbf{Case 2: Larger Search Scope.}
In Figure \ref{fig:rq4-2}, we compare the different behaviors of \autofl and \toolname in localizing the bug Jsoup-2~\footnote{\url{https://github.com/jhy/jsoup/issues/22}}. The issue is caused by the routine that greedily parses the data content not cleaning up the stack.
To fix this, the developers changed the \emph{parseStartTag} method within class \emph{org.jsoup.parser.Parser}.
In this case, we observe that the failure of \autofl originates from its selection of suspicious classes.
Specifically, during the first \emph{Function} call, \autofl assesses that \emph{``most likely the issue lies in either the Jsoup class or Document class''}, which results in the actual buggy class \emph{Parser} being overlooked.
Subsequently, as indicated in the third dialogue box on the left side of the figure, \autofl further requests to \emph{``get the list of methods in Document class that are covered by the failing test''}, exacerbating the error and ultimately preventing \autofl from identifying the expected buggy method.
In contrast, when referring to \toolname, we observe that although \emph{Module Knowledge} did not provide any useful context in this instance, semantic code search played a crucial role.
By focusing on software functionalities related to \emph{``data node''} and \emph{``script tag''}, \toolname successfully matched the buggy method \emph{parseStartTag} from the whole code base through code search, as the method \emph{``contains logic for handling data nodes within specific tags''}.


\subsection{Comparison with Learning-Based FL}

We also compared \toolname with various learning-based FL techniques, including FLUCCS~\cite{sohn2017fluccs} based on machine learning, DeepFL~\cite{li2019deepfl} based on deep learning, and GRACE~\cite{lou2021grace} based on graph neural networks.
Particularly, to align with practical application scenarios, we focused on the cross-project performance of these methods (i.e., trained on the Defects4J V1.2.0 benchmark~\cite{just2014defects4j} and tested on V2.0.0).

\begin{wraptable}[8]{r}{0.5\textwidth}
\centering
\caption{Results on Defects4J V2.0.0.}
\label{tab:dis}
 \resizebox{0.8\linewidth}{!}
{
\begin{tabular}{c|l|rrr}
\hline
\textbf{\# Bugs} & \textbf{Techniques}  & \textbf{Top1} & \textbf{Top3} & \textbf{Top5} \\ \hline
\multirow{4}{*}{280}  & FLUCCS      & 57   & 97   & 119  \\
& DeepFL & 43   & 89   & 112  \\
& GRACE  & 85   & 119  & 140  \\ \cline{2-5} 
& \cellcolor{tableblue} \toolname    & \cellcolor{tableblue} 105   & \cellcolor{tableblue} 137   & \cellcolor{tableblue} 178  \\ \hline
\end{tabular}
}
\end{wraptable}
As shown in Table \ref{tab:dis}, \toolname ranked 105 buggy methods in the first position, which is 48 more than FLUCCS, 43 more than DeepFL, and 20 more than GRACE.
On the Top3 and Top5 metrics, \toolname also consistently outperforms other FL approaches.
This indicates the substantial potential of employing semantic code search for fault localization.

\subsection{Threats to Validity}
{\bf Internal.}
The main internal threat comes from the data leakage problem.
As the Defects4J V1.2.0 benchmark~\cite{just2014defects4j} has been widely used for FL tasks since 2018, it may have been included in the LLM training corpus.
We mitigate this threat by evaluating the FL approaches on the newer version of Defects4J (V2.0.0).
Additionally, we collect 144 extra bugs from the GrowingBugs~\cite{jiang2022growingbugs} benchmark, which has almost never been used in previous FL studies.

{\bf External.}
Our evaluation of \toolname primarily focuses on software projects with Java unit tests.
Our findings may not be generalizable to other programming languages or different levels of testing.
To mitigate this threat, we adhered to the principle of scalability when implementing \toolname, and did not make any special designs for specific features of programming language syntax or bugs.
\section{Related Work}
\label{sec:related}

Fault localization (FL) is a crucial component of software debugging and has been extensively studied.
To help developers automatically find the fault locations in the entire code base, early FL techniques including Spectrum-based FL (SBFL)~\cite{abreu2006ochiai}, Mutation-based FL (MBFL)~\cite{wong2016survey}, and Information Retrieval-based FL (IRFL)~\cite{wang2015survey}.
These methods calculate the suspiciousness of fault locations through statistical analysis, without considering the semantics of the program itself.
With the advancement of artificial intelligence, machine learning and deep learning-based methods~\cite{li2019deepfl,lou2021grace} have emerged.
These methods embed code into vector space through feature engineering and representation learning, enabling neural networks to identify incoming bugs by fitting them with situations during training.
However, due to the lack of scalability and interpretability, these approaches struggle to be applied in practical scenarios~\cite{kochhar2016practitioner}.
Recently, the powerful code analysis capabilities of LLMs have offered new opportunities for FL, with preliminary attempts such as \autofl~\cite{kang2024autofl} and \agentfl~\cite{qin2024agentfl} have guided LLMs to autonomously navigate through the code base and identify defect locations.
In this paper, we propose \toolname which aims to boost FL from the perspective of semantic code search.
Notably, we differentiate \toolname from IRFL as the latter is mostly based on keyword-matching, which lacks the understanding towards semantic information.
\section{Conclusion}
In this work, we present \toolname, a novel Fault Localization (FL) approach inspired by semantic Code Search (CS).
\toolname proposes to regard FL tasks as a two-step process: the \emph{query generation} for generating a natural language query to describe the problematic software functionalities, and the \emph{fault retrieval} which employs CS to directly match buggy program elements from the entire code repository.
Evaluation results on 835 real bugs show that \toolname significantly outperforms other baseline approaches, potentially paving new pathways for the advancement of FL techniques.



\end{document}